\documentstyle[12pt,epsfig]{article}
\newcommand{\be}{\begin{equation}}
\newcommand{\ee}{\end{equation}}
\newcommand{\bq}{\begin{eqnarray}}
\newcommand{\eq}{\end{eqnarray}}

\newcommand{\bc}{\begin{center}}
\newcommand{\ec}{\end{center}}

\newcommand{\lt}{\tilde{\lambda}}

\newcommand{\dd}{\delta}

\newcommand {\s} {\sigma}

\def\(({\left(}
\def\)){\right)}
\def\[[{\left[}
\def\]]{\right]}
\def\bi{\bibitem}
\def \la{\langle}
\def\ra{\rangle}

\def\Jp{J_{i_1 \ldots i_p}^{l_1 \ldots l_p}}

\begin{document}

\title{{\bf Numerical study of a short-range $p$-spin glass model 
in three dimensions}}
\author{Matteo Campellone$^{*}$, Barbara Coluzzi$^{**}$ and
Giorgio Parisi$^{**}$}

\maketitle
\date{\today}

\begin{center}
(*) Departament de F\'{\i}sica Fonamental  \\ 
 Universitat De Barcelona \\
Diagonal 647, Barcelona, Spain\\
 (**) Universit\`a di Roma ``La Sapienza''\\
 Piazzale A. Moro 2, 00185 Rome (Italy)\\
 e-mail:{\it  campellone@roma1.infn.it, 
coluzzi@roma1.infn.it, giorgio.parisi@roma1.infn.it}\\

\vspace{2truecm}

\begin{abstract}
\noindent
In this work we study numerically a short range $p$-spin 
glass model in three dimensions. The behaviour of the model appears to be
remarkably different from mean field predictions. In fact it shares some 
features typical of models with full replica-symmetry breaking (FRSB).
Nevertheless, we believe that the transition that we study 
is intrinsically different
from the FRSB and basically due to non-perturbative contributions. 
We study both the statics and the dynamics of the system which seem to
confirm our conjectures.  
\end{abstract}

\end{center}
\vspace{1.cm}
\hspace{.1in} PACS Numbers 05-7510N

\section{Introduction}
\noindent
One of the presently unsolved theoretical issues concerning spin glasses is to
obtain a satisfying description of these systems in finite dimension.
By now, a quite complete picture has been achieved for spin glass models at 
mean field level \cite{mpv}. Quite generally, one finds a critical temperature 
$T_c$ which separates an ergodic phase from a low temperature spin glass 
phase in which ergodicity is broken. Below $T_c$ the spins are frozen in their 
local orientation, giving rise, because of the disordered nature of the 
interactions, to no total spontaneous magnetization. 

This transition is well described by a spontaneous replica symmetry breaking 
scheme \cite{mpv}. In the replica framework the order parameter of the spin 
glass transition is a function $q(x)$ defined on the real interval $[0,1]$ 
that changes from a high temperature constant form
 $q(x)=q_{0}$ to a low temperature function that can assume different values. 
The physical meaning of the function $q(x)$ is the 
following: at low enough temperatures the free energy landscape has many 
local minima separated by free energy barriers that grow, asymptotically,
as the size of the system. Below the critical temperature the system breaks 
the ergodicity and remains confined into one of these valleys. The values 
assumed by the order parameter function $q(x)$ are the possible values of the 
overlap $q_{\s \tau} = N^{-1} \sum_{i}^N \s_i \tau_i $ among two copies 
(replicas) of the system which are independently equilibrated in the same free 
energy landscape. 

If the two replicas freeze into the same pure state then the overlap will be 
the maximum allowed ($q_{EA}$) and it will provide a measure of the width of 
the state. The size of the interval where $q(x)=q$ gives the probability that 
the two replicas are found with an overlap $q$. More precisely, one can define
a quantity $P(q)=\frac{dx(q)}{dq}$ where $x(q)$ is the inverse function of 
$q(x)$. The function $P(q)$ is the probability distribution of the values of 
the overlap and it contains the same amount of information as the $q(x)$. 
At high temperatures one has a constant $q(x)=q_{0}$ which means that the 
$P(q)$ is a delta function on $q_0$.

Two main classes of spin glass models can be distinguished depending
on the character of their phase transition.

\begin{itemize}
\item
A first class of models undergoes, in mean field, to a full replica symmetry 
breaking (FRSB) transition in which the function $q(x)$ changes continuously
with the temperature.
The low temperature solution is a continuous function in which all values of 
the overlap between $q_{0}$ and $q_{EA}$ are attained. Correspondingly, the 
$P(q)$ will change to a non-zero continuous function between $q_{0}$ and 
$q_{EA}$ with an additional delta function on $q_{EA}$.

This transition is continuous in the sense that $q_{EA}-q_{0}$ is small for 
$T$ close to $T_c^-$.

These models describe with success the physics of real spin glasses. This kind
of transition is second order. At a critical temperature $T_{c}$ the 
spin glass susceptibility diverges and long range correlations establish.

\item
A second class of spin glasses has a discontinuous low temperature function 
$q(x)$ that can assume a finite number of values. Usually $q(x)$ assumes only 
two values $q_0$ and $q_1$($=q_{EA}$) and the transition involves 
one step of replica symmetry breaking (1RSB). In most cases the 
transition is discontinuous in the sense that $q_{0} - q_{EA} \simeq O(1)$ 
as soon as $T < T_c$. 

The low temperature $P(q)$ can be written in the following
form $$P(q) = m \dd(q_0) + (1-m)\dd(q_1)$$
where $m$ is a parameter which is smaller than unity and it decreases with 
the temperature. 

In this second class of models the distribution $P(q)$ is zero
 between $q_0$ and 
$q_1$ which are the only allowed values for the overlap.
 This is very different from the case of models of the 
first class where the low-temperature $P(q)$ shows a plateau from zero to 
$q_{EA}$ and there is no $\dd(q_0)$.
 In absence of 
magnetic field one always has $q_0 = 0$.
 We shall stick to this case in the following.

\end{itemize}

Another important feature of mean field 
spin glasses is that the dynamical equations
for the two-time correlation 
functions show a singularity at a critical temperature 
$T_d \ge T_c$. It turns out that for models of the first class the dynamical 
and statical critical temperatures coincide,
while in models of the second class one has $T_d > T_c$.
The existence of a dynamical singularity at a 
higher temperature than the static critical temperature is an aspect of 
these models which is shared by structural glasses \cite{kitiwo}.

The mode-coupling (MC) theory describes quite well 
the dynamics of supercooled liquids \cite{gotze} approaching their 
structural glass transition. The MC dynamical equations predict a 
singularity at a temperature $T_{mc}$ which is higher than the glass 
temperature $T_g$ predicted for real glasses in a static Adam-Gibbs scenario. 

There has been recent 
effort to approach the structural glass transition by means of the techniques 
developed for disordered systems such as spin glasses \cite{pa1,pa2}.
An effective disorder is in fact induced in structural glasses 
by the slow dynamics of the microscopic variables. 

So, despite the absence of quenched disorder in the Hamiltonian and the
possible presence of a crystalline groundstate (which however corresponds to
a golf-shaped minimum in the free-energy landscape and can therefore been
considered irrelevant to the physics of the system), it is believed
 that the behaviour of structural glasses
is very similar to the one of 1RSB mean field spin glass models.
Recent numerical results seem to support this hypothesis \cite{pa3}-
\cite{cafrpa2}. 
Of course, in finite dimensional glasses, the effect of the decay of 
metastable states by activated processes have to be taken into account,
 and are an element of difference with respect to mean field spin glasses.  
An interesting issue is to generalize long range 
models of spin glasses with 1RSB to models with 
short range interactions to see if the finite dimensional effects 
are similar to the case of real glasses or not. 

In this work we present a numerical study of a $3d$ short-ranged 
version of the $p$-spin model \cite{capara,frpa,maetal,papiri}. 
The model is provided of a parameter $M$ 
that ensures the mean field solution to be exact when $M$ is sufficiently 
large.
 A perturbative analysis around the mean field solution predicts a 
behaviour qualitatively similar to the MF \cite{capara}. Nevertheless, 
numerical simulations give evidence for a diverging correlation length
when approaching the critical temperature from above \cite{frpa}. 

We study the model in $d=3$ for $p=4$ and $M=3$.
We look at the behaviour of the function $P(q)$ 
which appears definitely non trivial at low temperatures. 
The form of the $P(q)$ that we find seems to be relevantly 
different from the shape that one has in the case of the mean field $p$-spin 
model. Our results indicate the occurring of a continuous transition.
 Furthermore, a numerical study of the dynamics of this
model seems to provide evidence for a dynamical transition at the same 
critical temperature of the static one as in the cases of models with a 
continuous transition.
Nevertheless we believe that this transition is very peculiar and is not
due to an instability of the saddle point as in the case of models with FRSB.
We address the peculiar nature of this transition 
to the formations of large regions
of highly correlated spins which are localized by the quenched disorder 
and whose typical size grows as $T \to T_c$. 
Before presenting the short range model and our numerical results, we will 
devote the following section to describing the known results 
for the 1RSB class of mean field spin glasses and to do so we will focus 
on a particularly simple model belonging to this class.

\section{The long-range $p$-spin glass model}
\noindent
In this section we shall briefly resume the known results for
a long range model that has been extensively studied in the past and,
because of its 
simplicity, is a recommendable 
paradigm for mean field spin glasses undergoing a 1RSB transition.

The model in zero magnetic field is defined by the $p$-spin interaction 
Hamiltonian
\be
{\cal H}_{p}(\{ \s \}) =
 -\sum_{(1\leq i_{1}<i_{2}<\cdots<i_{p}\leq N)}\hspace{-.5cm}
J_{i_{1},i_{2},\cdots,i_{p}} \s_{1}\cdots \s_{p}.
\label{Hpspin}
\ee

The spin variables can be either Ising spins ($\s_i = \pm 1$) or continuous 
soft spin variables which have to be constrained by some normalizing 
condition such as

\be
\sum_{i=1}^{N} {\s_{i}}^2 = N,
\label{sph}
\ee
where the interactions 
$J_{i_{1},i_{2},\cdots,i_{p}}$ are random variables 
distributed with
\be
P(J_{i_{1},i_{2},\cdots,i_{p}}) = \left[\frac{N^{p-1}}{\pi J^2
p!}\right]^{-\frac{1}{2}} \exp{\left[-\frac{(J_{i_{1},i_{2},\cdots,i_{p}})^2
N^{p-1}}{J^2 p!}\right]}.
\label{distrprobp}
\ee

The scaling of the variance with $N^{p-1}$ ensures the free energy to be
extensive. The mean field solution of the model is exact when the limit 
$N \to \infty$ is taken since the interactions are long ranged. The physics 
of the Ising and of the spherical $p$-spin glass model is essentially the same 
if $p>2$. Here we will review some well established results in the spherical 
case which can be exactly solved.
For $p=2$ this model does not belong to the 1RSB class, and exhibits 
a continuous transition.

The model presents a 1RSB discontinuous transition at a temperature 
$T_c$. One can distinguish two other relevant temperatures $T_d$ and $T_m$, 
where $$T_c < T_d < T_m.$$

The physics of the model is characterized by the appearance of a large 
number of solutions different from the paramagnetic one. We can schematically 
resume the behaviour of the model with respect to temperature as follows

\begin{itemize}

\item
For $T>T_m$ the model is in a paramagnetic phase which is the only minimum 
of the free energy.

\item
For $T_d < T < T_m$ the free energy landscape changes. The physics of the 
model is still dominated by the paramagnetic state, but other local minima 
of the free energy appear. These minima have a lower energy than the 
paramagnetic state but their 
free energy is higher and therefore they are still irrelevant in the 
thermodynamic limit. There is an exponential number of these minima. One can 
define the complexity $\Sigma$ as the logarithm of the number of states that 
becomes extensive below $T_m$.

The free energy of the ensemble of these states is therefore
\be
{\cal F}_s = E_s - T S - T \Sigma 
\ee

Above $T_d$ one has $F_s > F_{p}$ where $F_{p}$ is the free energy of the 
paramagnetic state. So in this regime the dominant solution is still the 
paramagnetic one.

\item
In the whole range of temperatures such that $T_c < T < T_d$ one finds that
the free energy of the paramagnetic state equals the free energy of the 
ensemble of local minima {\i.e.} $${\cal F}_s = F_{p}. $$

It turns out that at the temperature $T_d$ the solutions to the dynamical 
mean field equations change. In this sense $T_d$ is the dynamical critical
temperature. This corresponds to the fact that, if the system starts from a 
random initial configuration, the system will eventually fall into a 
metastable state where it will be trapped for an infinite time.

\item
For $T$ close to $T_c$, $\Sigma(T)$ vanishes linearly with $T-T_c$. At
$T=T_c$ the number of local minima becomes non-extensive and a static 
transition of the 1RSB kind occurs.
\end{itemize}

As the system is cooled below $T_d$ a dynamical transition occurs. One can 
easily write some closed equation for two time quantities such as 

\bq
C(t+t_{w},t_{w}) & = & \frac{1}{N} \sum_{i}^{N}\la \s_{i}(t+t_{w})
\s_{i}(t_{w}) \ra, \\
G(t+t_{w},t_{w})& = & \frac{1}{N} \sum_{i}^{N}
 {\partial\la \s_i(t+t_{w} )\ra \over \partial h_i(t_{w} )}.
\label{corrs}
\eq

If the system has reached equilibrium, the two time quantities depend only 
on the differences between the two times and are related by the equation

\be
G(t) = -\frac{1}{T} {\partial C(t) \over \partial t}
\theta(t). 
\label{fdt1}
\ee

These two properties, time-translation invariance (TTI) and the 
fluctuation-dissipation theorem (FDT) are properties of equilibrium dynamics.

Below $T_d$ the system has a regime of slow dynamics where, for large time 
differences $t \simeq t_w$, the relaxation of C and G is slower the 
larger $t_w$ is {\em i.e.} the {\em older} the system is.
In this regime one can write the two-time quantities under the form

\be
C(t+t_w,t_w) = \tilde{C} \(( \frac{h(t+t_w)}{h(t_w)} \)) \hspace{1cm}
G(t+t_w,t_w) =\frac{1}{t+t_w} \tilde{G}(\frac{h(t+t_w)}{h(t_w)}),
\ee

Relation (\ref{fdt1}) is not valid under $T_d$, but a generalized FDT can
be written introducing a function $X_{t_w}(C)$ \cite{cuku}
by the relation

\be
G(t) = -\frac{X_{t_w}(C)}{T} {\partial C(t) \over \partial t}
\theta(t). 
\label{ofdr}
\ee

For large waiting times one has that $X_{t_w}(C)\rightarrow X(C)$.
The function $X(C)$ generally characterizes the type of aging dynamics of 
the model. In models where $T_d=T_c$ ({\em i.e} $p=2$) 
one has that $X(C) = x(q)$ while this equation does not hold in general
though the two functions remain equal in structure \cite{cuku2}.
The dynamical transition in mean field spin glasses is interpreted as
due to the freezing of the system into one among the $O(\exp{N})$
metastable states, which for a long range model have an
infinite lifetime in the $N \to \infty$ limit. 

The presence of metastable states with infinite time life is just an artifact
of the mean field approximation and we do not expect $T_d$
 to mark an effective 
dynamical transition in a finite dimensional model. Correspondingly
one should find \cite{frmepape} that the dynamical function $X_{t_w}(C)$ 
becomes equal to the static $x(q)=\int_0^{q} dq' P(q')$ in the limit of
$t$, $t_w \rightarrow \infty$ (in which $C \rightarrow q$) also for
1RSB models. Nevertheless,
it should be still possible to observe a $T_d$ higher then the critical 
temperature $T_c$. In a finite-dimensional system that has a 1RSB mean field
solution $T_d$ should mark 
the onset of a two step relaxation process: 
the mean field-like relaxation of the system within 
a metastable 
state and the following decay of the metastable 
state due to activated processes.

\section{The model}
\noindent
The model that we study in this work is a short range version of the 
Ising $p$-spin glass. We believe that for our purposes the Ising
and spherical version of the $p$-spin are essentially equivalent. 

The model is defined as follows. We consider a cubic tridimensional lattice  
of side $L$. On each site of the lattice we put $M$ Ising spins.
The total number of spins is therefore $N=M \: L^d$.
Each spin interacts with $M-1$ spins on its same site and with
$2dM$ spins on its nearest-neighbour sites. The interactions $\Jp$ are 
quenched random variables.

The Hamiltonian of the model can be written as 
\be
{\cal H}_{p}(\{ \sigma \}) = \sum_{<i_1,\cdots,i_p>}^{L^d}
 \sum_{l_1,\cdots,l_p=1}^M \Jp {\sigma_{i_1}}^{l_1}\cdots 
{\sigma_{i_p}}^{l_p}.
\ee

By $\sum_{<i_1,\cdots,i_p>}^{L^{d}}$ we mean the sum over all the sites of 
the lattice 
taking, for each couple of adjacent sites $i$ and $j$, $p-k$ of the 
$i_1,\cdots,i_p$ indexes equal to $i$ and $k$ indexes equal to $j$. 
In other words, for each nearest neighbour sites $i$ and $j$, every 
interaction involves $p-k$ spins of site $i$ and $k$ spins of site $j$ 
with $k$ running from zero to $p$.

We consider discrete $(\pm 1)$ spin variables and we call 
${\sigma_{i_r}}^{l_r}$ the $l_r^{th}$ spin of site $i_r$ with 
$l_{r}$ running from $1$ to $M$. 

For large $M$, each spin interacts with a large number of nearest neighbours
and, in the limit $M \to \infty$, the mean field approximation has to be 
exact. An analytical study of the model for large $M$ has been performed in 
\cite{capara} where a Gaussian $P(\Jp)$ is considered. 
The mean field solution of the model coincides with the solution 
of the long range Ising $p$-spin.

One could hope to peep into the finite-$M$ case by perturbing around the 
$M \to \infty$ limit. Calculating the $O(1/M)$ corrections to the free 
energy and the Gaussian propagators one sees that a perturbative approach
shows a transition which is very similar to the MF one, with a 
discontinuous $q(x)$ and no diverging low-momentum static propagators.
In the model that we simulated we chose the following distribution for the
couplings
 
\be
P(J) = \frac{1}{2} \dd_{J,1} +\frac{1}{2} \dd_{J,-1}. 
\ee

The fact that the distribution of the couplings is non-Gaussian 
should not be source of relevant differences from the analytical calculation.
Any result in disagreement with the results of \cite{capara} should 
be imputed to finite dimensional effects which can not be reached with 
$O(1/M)$ calculations.  

Here we take $M=3$ and $p$=4. Since $p$ is even, the Hamiltonian is  
invariant for inversion of all the spins ( $\{ {\sigma_i}^l \} \rightarrow \{ - 
{\sigma_i}^l \}$ ) and the $P(q)$ is correspondingly symmetric, i.e. 
$P(q)=P(-q)$. 

As usual in spin glass simulations we consider two replicas (with the same 
configuration of disorder) of the system, with spin $\{ \sigma \}$ and 
$\{ \tau \}$ respectively, that evolve simultaneously and independently.
We define
\begin{equation}
Q \equiv \frac{1}{N} \sum_{i=1}^{L^3} Q_i \equiv 
\frac{1}{N} \sum_{i=1}^{L^3} \frac{1}{M} \sum_{l=1}^{M} {\sigma_i}^l
{\tau_i}^l
\end{equation}
The order parameter probability distribution $P(q)$ is then given by
\begin{equation}
P(q) \equiv \overline{ \langle \delta ( q - Q ) \rangle }
\label{pdiq}
\end{equation}
where $\langle \cdot \rangle$ means thermal average and 
$\overline { ( \cdot ) }$ means average over disorder.

When looking at the out of equilibrium behaviour of the system it is 
useful to consider \cite{frari} the staggered magnetization
\begin{equation}
m_s[h](t)= \frac{1}{L^3} \sum_{i=1}^{L^3} \langle 
\frac{1}{M} \sum_{l=1}^{M} \sigma_i^l(t) \rangle,
\end {equation}
that is related to the response function $G(t_1,t_2)$ by
\begin{equation}
G(t_1,t_2)= \frac{\delta m_s[h](t_1)}{\delta h(t_2)}.
\end{equation}

One has therefore
\begin{equation}
m_s[h](t_1) = \int_{-\infty}^{t_1} dt_2 \frac{\delta m_s[h](t_1)}
{\delta h(t_2)} =  
\int_{-\infty}^{t_1} dt_2 G(t_1,t_2) h(t_2) + {\cal O} ( h^2 ), 
\end{equation}
which is just the linear-response theorem.

By applying the generalized FDT relation (\ref{ofdr}) one obtains:
\begin{equation}
m[h](t_1) \simeq  \beta \int_{-\infty}^{t_1} dt' X[C(t_1,t_2)] 
\frac{\partial C(t_1,t_2)}{\partial t_2} h(t_2).
\end{equation}

If the small perturbing magnetic field is turned on at the time $t_w$ to a 
constant time-independent value $h_0$ we get
\begin{equation}
\begin{array}{lclcl}
m_s[h](t_1) & \simeq & \beta h_0 \int_{t_w}^{t_1} dt' X[C(t_1,t_2)]
\frac{\textstyle \partial C(t_1,t_2)}{\textstyle \partial t_2} & = & \\ 
& & & &\\
& = & \beta h_0 \int_{C(t_1,t_w)}^{1}
du X[u] & = &\beta h_0 S(C),
\end{array}
\end{equation}
where we have used $C(t_1,t_2)$=1 for $t_1=t_2$.

In the case of finite dimensional models, for large $t_w$, $t_1=t+t_w$, 
one has \cite{frmepape} $C \rightarrow q$, $X(C) \rightarrow x(q)$, where
$x(q)=\int_0^q dq' P(q')$. This means that the staggered magnetization measured
by out of equilibrium simulations is related to static quantities 
and more precisely
\begin{equation}
S(C) \rightarrow y(q)=\int_q^1 dq' x(q') \hspace{.3in} 
\end{equation}
We are therefore allowed to confirm results on the equilibrium probability
distribution of the overlaps $P(q)$ by looking at the dynamical behaviour
of large systems.

\section{On the behaviour at the equilibrium}

\subsection{Simulations}
\noindent
In order to measure equilibrium quantities we use Parallel Tempering (PT)
\cite{tereorwh}. We consider size ranging from $L$=3 up to $L$=6 ($N$=648),
simulating contemporaneously two independent sets of $n$ replicas that move
between $ \beta_{min}$= 0.14 and 
$\beta_{max}$= $\beta_{min}+n \Delta \beta$=0.5 (i.e. down to $T_{min}$=2.0),
where $\Delta \beta$=0.4 for $L$=3, 4 and $\Delta \beta$= 0.2 for $L$=5, 6.
We perform $2^{18}$ $PT$ steps for $L$=3, $2^{19}$ for $L$=4 and $2^{21}$ for
$L$=5, 6, that means up to 2 millions of MC steps for each of the 
$2 \: n$=38 replicas.  

\begin{figure}[htbp]
\begin{center}
\leavevmode
\epsfig{figure=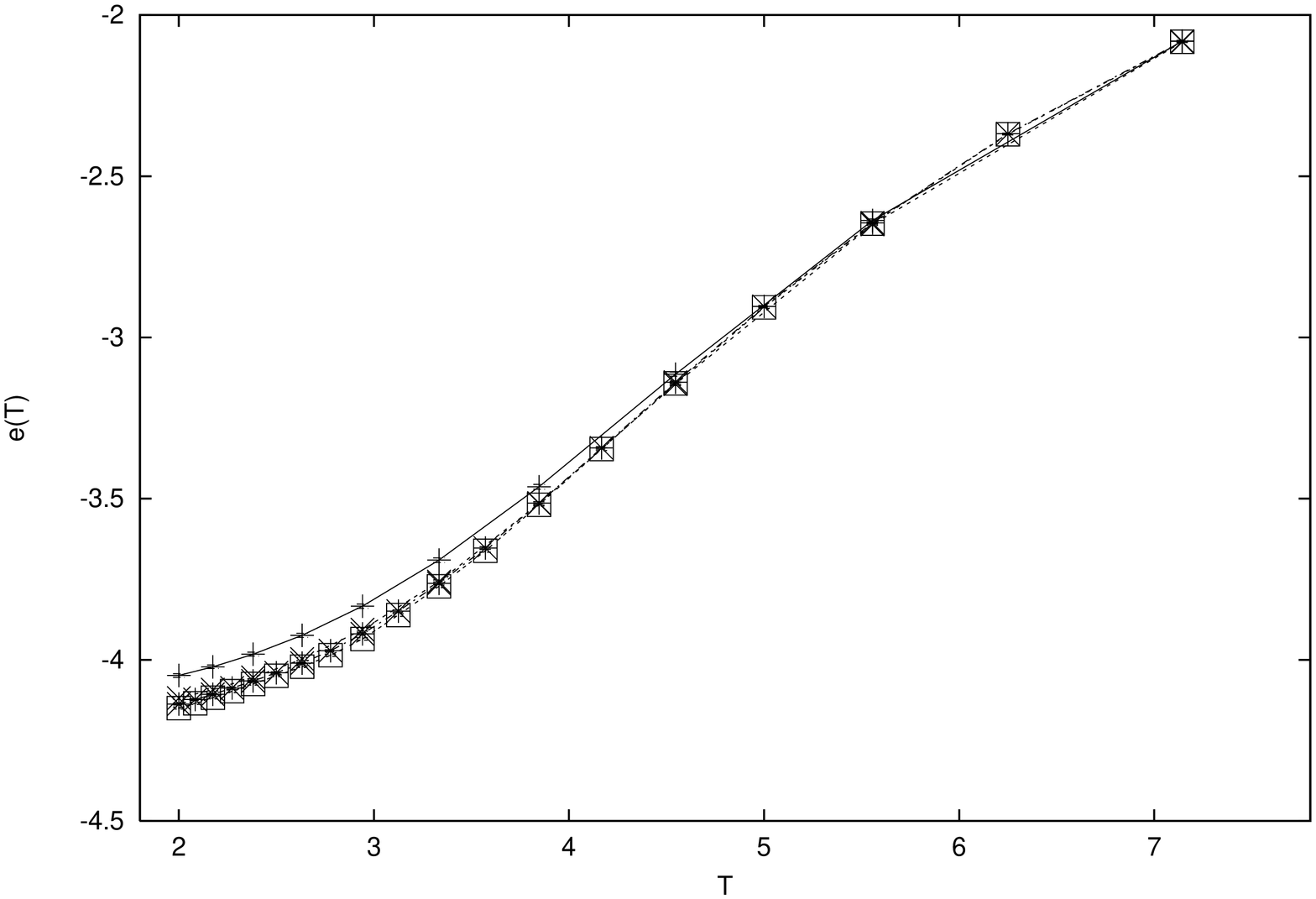,angle=270,width=6cm} 
\epsfig{figure=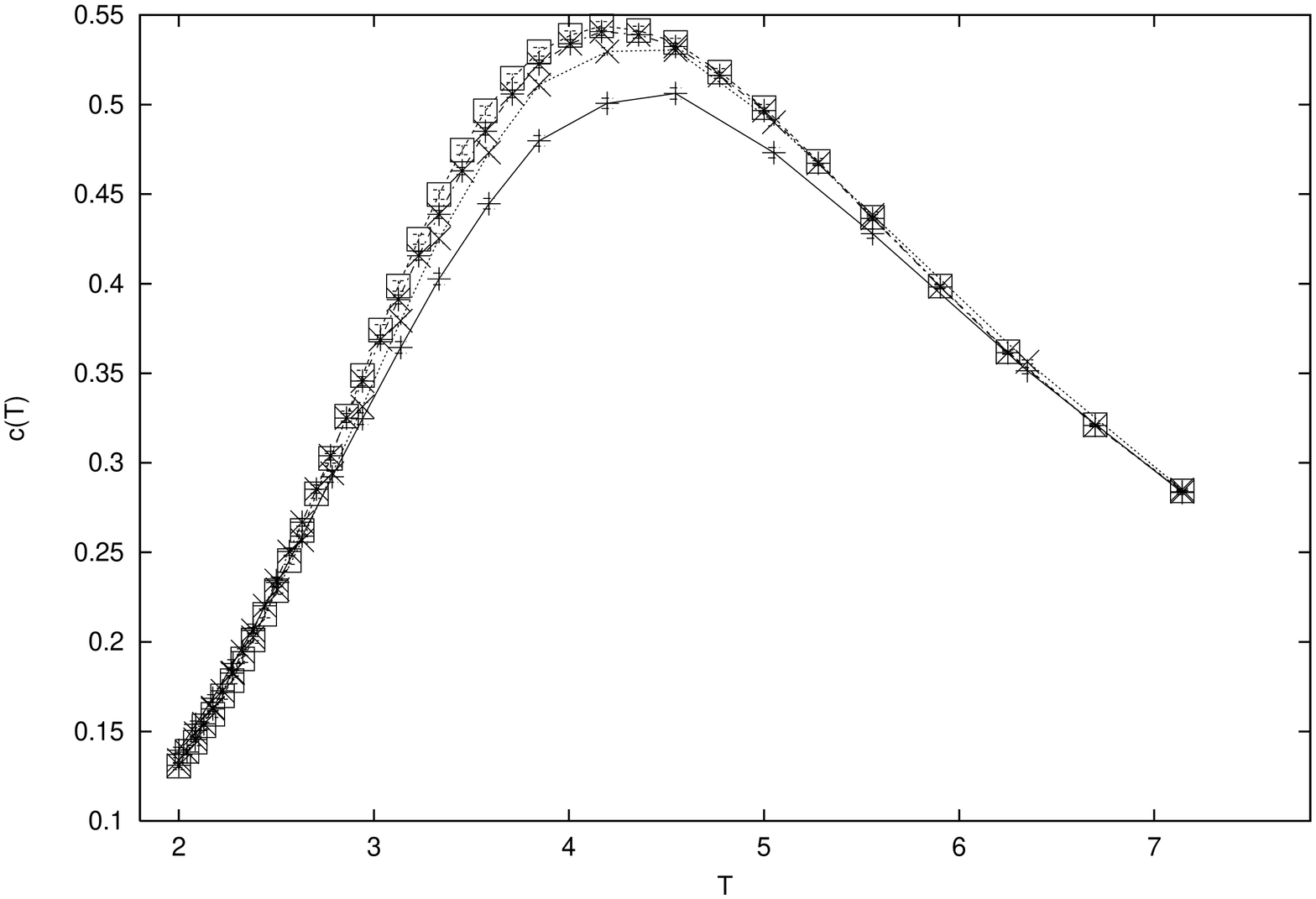,angle=270,width=6cm}
\end{center}
\caption{Data on the energy density $e$ (left) and on the specific heath $c$ 
(right) as a function of $T$ for $L$=3(+), $4 (\times)$, $5 (\ast)$ and 
$6 (\Box)$. Lines are only to join neighbouring points.}
\label{enecs}
\end{figure}

Thermalization is checked in different ways:
\begin{itemize}
\item We measure all the relevant quantities during
the whole run (in each $2^t$ - $2^{t+1}$ interval), looking at the convergence 
to equilibrium and checking that there are no shifts of the mean values in the
last intervals. In particular, we do not find evident changes 
in the behaviour of $P(q)$. The presented data are the ones collected in the 
last half of the run. 
\item We check that each replica moves several times from an extrema of 
the temperature range to the other and back. 
\item We evaluate the specific heath $c$ both using $ c= 
\partial < e > / \partial T$ and using $T^2 \: c = < e^2 > - < e >^2$, 
checking the compatibility of the results (see [Fig. \ref{enecs}]).
\end{itemize}

Errors are estimated from the fluctuations between different disorder
realizations. We consider 400 samples for $L$=3, 300 for $L$=4, 
240 for $L$=5 and more than 100 for $L$=6.

\subsection{Results on $\chi_{SG}$ and $P(q)$}
\noindent
We present in [Fig. 1] data on the energy and the specific heath as 
a function of $T$ for the different sizes considered. Finite size effects
are quite evident for $L$=3 while the observed behaviours result very similar
for larger $L$ values.

The spin glass susceptibility $\chi_{SG} \equiv M \: L^3 \langle q^2 \rangle$ 
is plotted in [Fig. 2] for the different sizes considered. 
The obtained behaviour confirm \cite{frpa} the presence of a non-zero 
$T_c$ thermodynamical 
transition which appears to be of second order. Data agree well 
[Fig. 3] with the finite size scaling law

\begin{equation}
\chi_{SG}(L,T)/L^{2-\eta}=\tilde{\chi}_{SG} \left( L^{1/ \nu} (T-T_{c}) 
\right), 
\end{equation}
where we use $T_c \simeq 2.6$, $\eta \simeq 0$ and $\nu \simeq 1$. The value 
of $\nu$ results slightly higher than the previously estimated 
\cite{frpa} $\nu \simeq 2/3$, this being probably due to some corrections to
finite size scaling.

\begin{figure}[htbp]
\begin{center}
\leavevmode
\centerline{\epsfig{figure=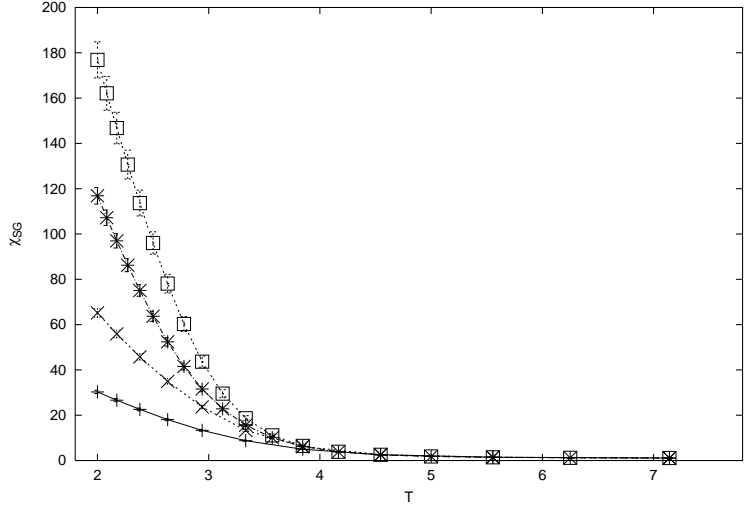,angle=270,width=10cm}} 
\caption{Data on the spin glass susceptibility ${\chi}_{SG}$ as a 
function of $T$ for 
$L$=3(+), $4 (\times)$, $5 (\ast)$ and $6 (\Box)$.
Lines are only to join neighbouring points.}
\end{center}
\label{su}
\end{figure}

\begin{figure}[htbp]
\begin{center}
\leavevmode
\centerline{\epsfig{figure=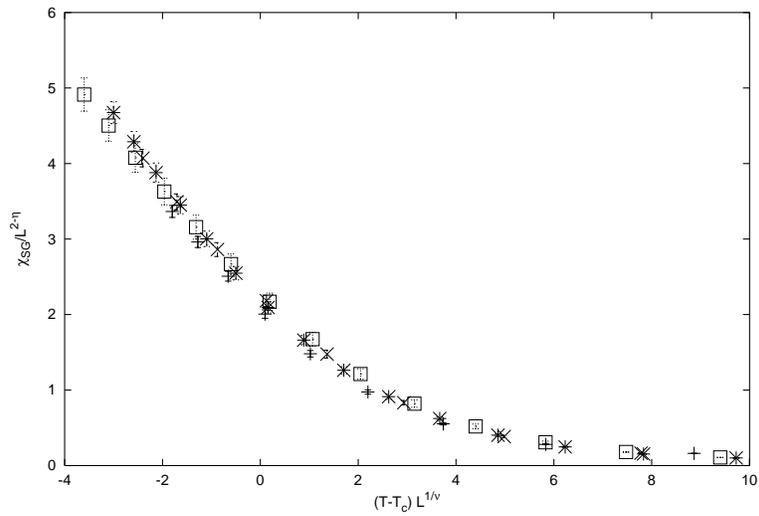,angle=270,width=10cm}} 
\caption{Data on the spin glass susceptibility ${\chi}_{SG}/L^{2}$ as a 
function of $(T-2.6) \: L$ for 
$L$=3(+), $4 (\times)$, $5 (\ast)$ and $6 (\Box)$.}
\end{center}
\label{suscala}
\end{figure}

In [Fig. 4] we plot $P(q)$ for $L$=5 at different 
temperatures, from $T$=3.33 down to $T$=2.0. The behaviour emerging from
these data results quite puzzling.

When approaching $T_c$, $P(q)$ becomes clearly non Gaussian but there is no 
evidence for the transition being discontinuous in the order 
parameter. Our data are compatible both with $q_1$=$O(T-T_c)$ at 
$T \tilde{<} T_c$ and with the possibility of a finite $q_1$=$O(1)$, 
hidden by finite size effect. 

At lower temperatures the situation appears still less clear: From the mean 
field theory we expect only two possible values of the overlap, i.e. $q_0=0$ 
and $q_1$ which increases when lowering $T$, whereas we find a $P(q)$ which is 
clearly greater than zero in all the range $[-q_1,q_1]$ also at the lowest 
temperature considered $T$=2.0. 
Moreover we observe, as expected, the weight of the peak in $q_0$ to decrease 
when lowering $T$ but it widens, although never disappear completely.

We could suppose that these are finite size effects. It should also 
be noted that in the mean field spherical $p$-spin all the values of the 
overlap are allowed within metastable states \cite{cagipa} and the contribute 
of these states is expected to be negligible only at very large $N$. 
Nevertheless these effects do not appear to decrease as $L$ increases.

\begin{figure}[htbp]
\begin{center}
\leavevmode
\centerline{\epsfig{figure=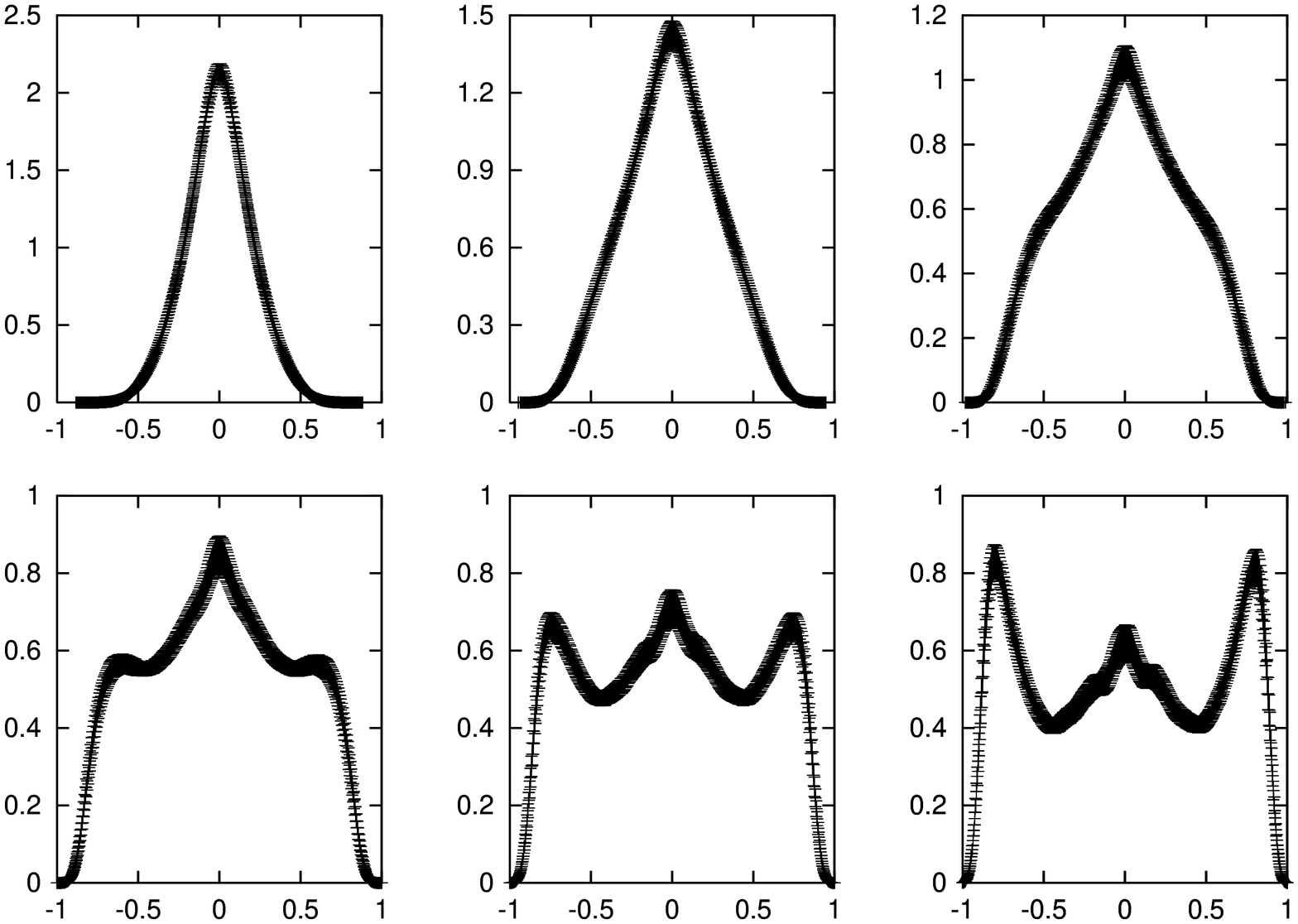,angle=270,width=10cm}} 
\caption{Data on $P(q)$ for $L$=5 (240 samples) at different temperatures. 
From left to right and top to bottom 
$T$=3.33, 2.94, 2.63, 2.38, 2.17 and 2.0.} 
\end{center}
\label{pql5}
\end{figure}

\begin{figure}[htbp]
\begin{center}
\leavevmode
\centerline{\epsfig{figure=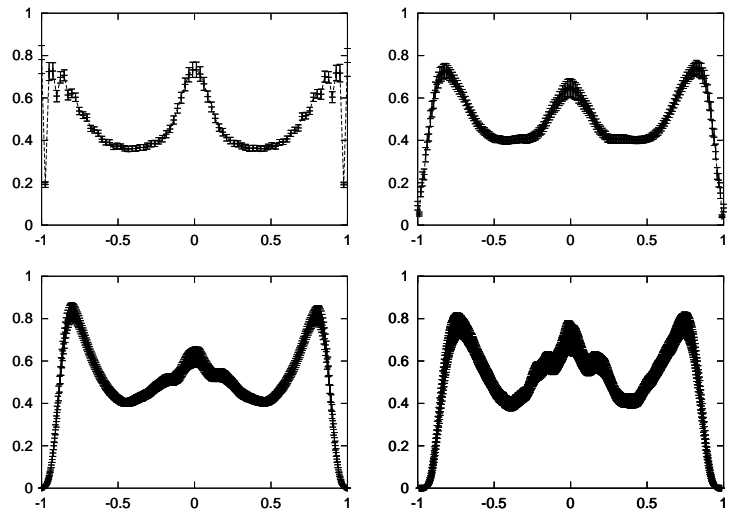,angle=270,width=10cm}} 
\caption{Data on $P(q)$ for the different sizes. 
From left to right and top to bottom 
$L$=3, 4, 5 and 6.}
\end{center}
\label{pqvl}
\end{figure}

\begin{figure}[htbp]
\begin{center}
\leavevmode
\epsfig{figure=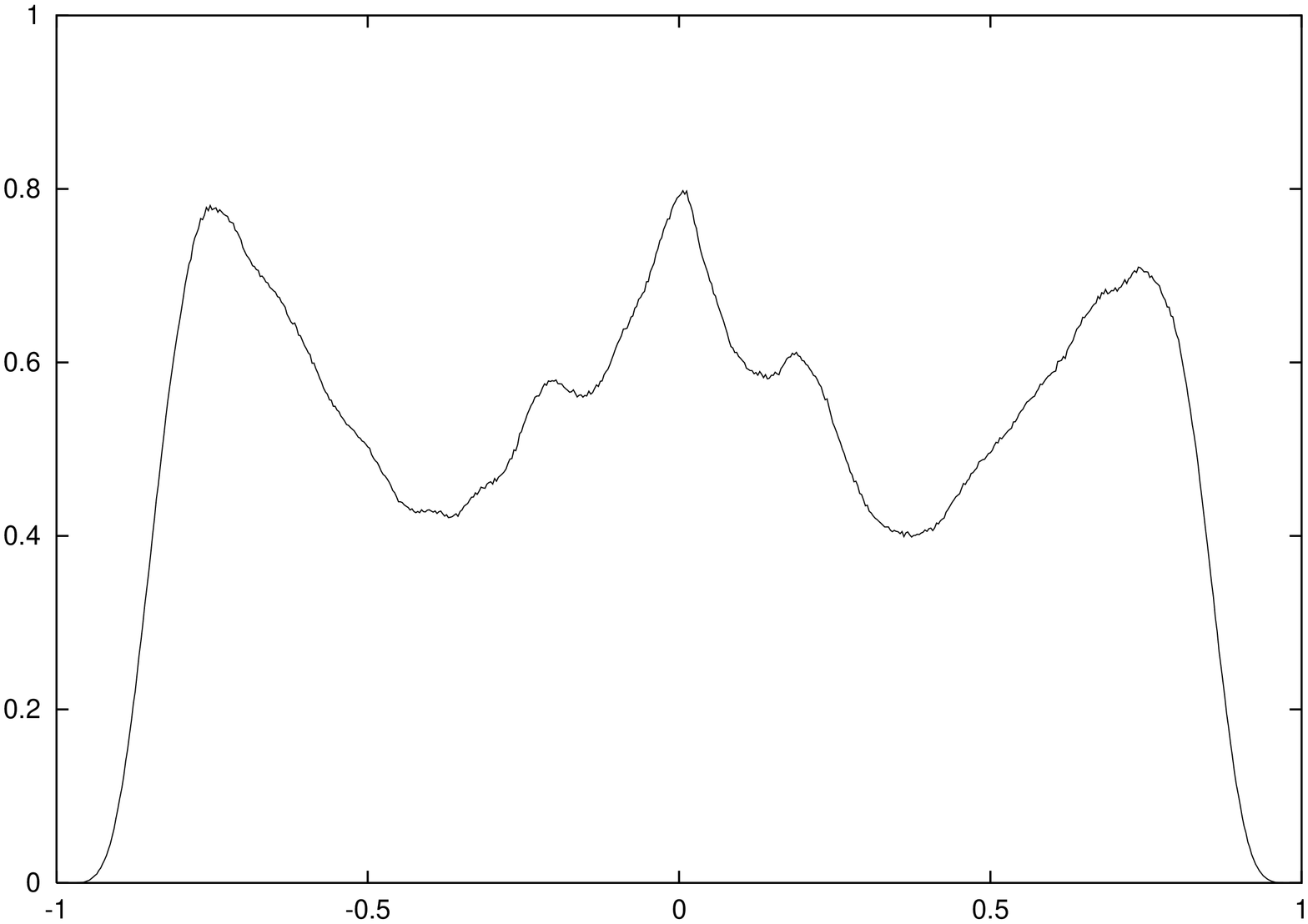,angle=270,width=5cm} 
\epsfig{figure=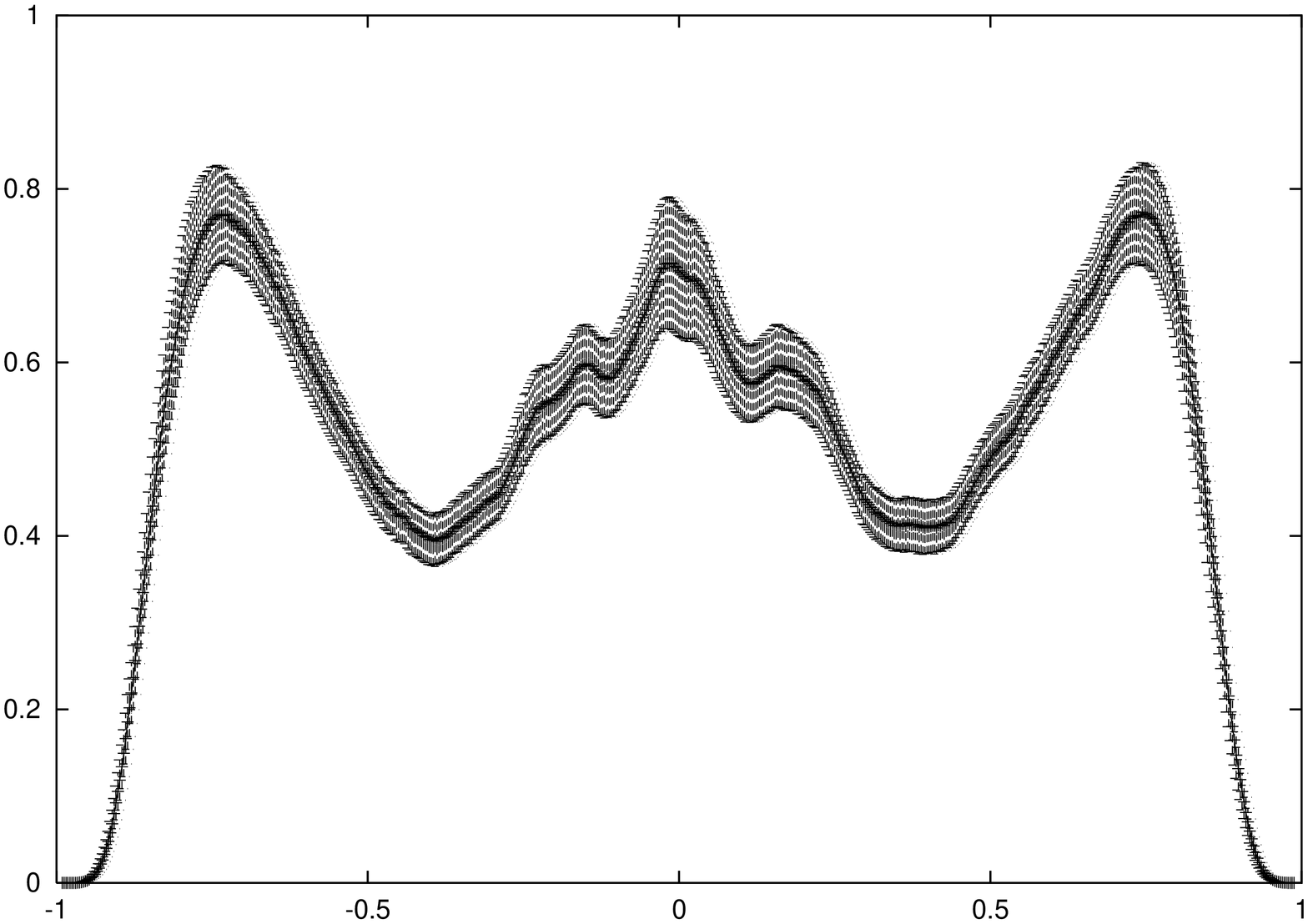,angle=270,width=5cm}
\caption{Data on $P(q)$ for $L$=6 (118 samples) at the lowest temperature 
considered $T=2.0$ as obtained in the last half of the run (right) and in the
previous interval (i.e. in the previous $2^{20}-2^{19}$ PT steps).} 
\end{center}
\label{pql6}
\end{figure}

As it can be seen from [Fig. 5], where we present data on
$P(q)$ at $T$=2.0 for the considered sizes, the minimum value $P_{min}= 
\min_{q \in [-q_1,q_1]} P(q)$ for $L$=3 ($N$=81) is compatible within the
errors with the one for $L$=6 ($N$=648). 
Moreover the broadening of the
peak in $q_0$ when considering larger sizes is very evident if we
compare data for $L$=3 with the ones for $L$=5. 

The comparison between the $L$=5 and $L$=6 data on $P(q)$ results puzzling 
since the weight of the peak in $q_0$ seems to be larger in the last case.
On the other hand it is well known that a correct estimation of 
$P(0)$ in spin glass simulations requires great care since the possible 
presence of non well thermalized samples causes the overestimation of this
value.
 
In order to clarify this point we plot in [Fig. 6] $P(q)$ for
$L$=6 at $T$=2.0 in the last half of the run (the same data that in
[Fig. 5]) and in the previous interval (i.e. in the previous 
$2^{20}-2^{19}$ PT steps). Although data are compatible within the errors, 
the shift of the peak in $q_0$ is quite evident (this is an effect that we do 
not observe in $L$=5 data) and it shows that some of the samples are still 
not perfectly equilibrated at the beginning of the last half of the run.
We are therefore suspicious that $2^{23}$ PT steps would be needed to 
be sure of achieving a complete thermalization for $L$=6 down to $T$=2.0, 
which means unfortunately too much CPU time.

If the peak in $q_0$ would disappear completely when considering larger
size, the $P(q)$ would display a FRSB-like 
behaviour at low temperatures, i.e. two peaks in $\pm q_1$ separated by a 
continuous $plateau$. Nevertheless, by looking carefully both to $L$=5 and 
$L$=6 data at low temperatures and to the previous [Fig. 5], 
we note that the broadening of this peak is related to the appearance of 
smaller, not well pronounced peaks near $q_0$ and that the behaviour of the
$P(q)$ near and below $T_c$ results very different from
the one usually encountered in short range models with FRSB \cite{cipari}. 

\begin{figure}[htbp]
\begin{center}
\leavevmode
\centerline{\epsfig{figure=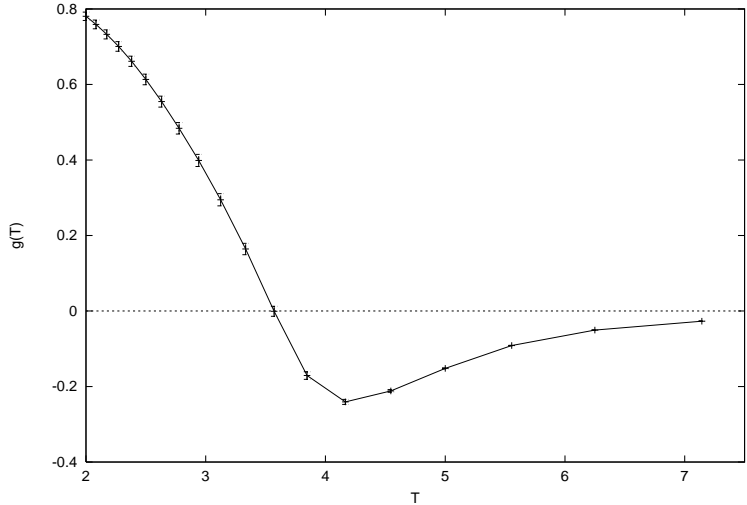,angle=270,width=10cm}} 
\caption{Data on $g(T)$ for $L$=5 (240 samples) at different temperatures. 
From left to right and top to bottom 
$T$=3.33, 2.94, 2.63, 2.38, 2.17 and 2.0.} 
\end{center}
\label{cum}
\end{figure}

To emphasize the last observation, we present in [Fig. 7]
data on the cumulant
\begin{equation}
g(T) = \frac{1}{2} \left ( 3 - 
\frac{\overline{\langle q^4 \rangle}}{\overline {\langle q^2 \rangle}^2}
\right ).
\end{equation} 
This is the well known dimensionless parameter usually successful for
evaluating $T_c$, since $g(T)$ should go to zero above $T_c$ in the 
thermodynamic limit and the curves for different values of $N$ should
cross at the critical temperature because of finite size scaling. In the 
FRSB mean field
models
as well as in their short range versions (EA models) $g(T)$ for finite $N$ 
is an $always~positive$ function of $T$ that increases quite regularly 
when lowering the temperature. The behaviour displayed in [Fig. 7] 
results therefore in evident disagreement with the one expected in the 
usual FRSB case: we find a $g(T)$ which is a non monotonic 
function of $T$, negative on a large part of the range of temperatures 
considered.
This accounts for the fact that the function $P(q)$ develops non-Gaussian
 tails in a range of temperatures above $T_c$ detecting the presence of 
non perturbative effects (see the related discussion in the conclusions).  
 The shown data correspond to $L$=5 but the behaviour is 
qualitatively similar for the other $L$ values that we have considered and it 
suggests that 
this model belongs to a different universality class from the EA 
spin glass one.

In conclusion, more extensive simulations are necessary to achieve a better 
understanding of the short range $p$-spin glass. Nevertheless it 
seems quite evident from our data that we are looking at a model characterized
by a $P(q)$ that is in disagreement with the one of 1RSB mean field models 
but that appears also very different from the one usually encountered in FRSB 
models. This is a quite new 
result that may have interesting physical interpretations as we will discuss 
in the following.

\section{On the dynamical behaviour}

\subsection{The approach to the equilibrium}
\noindent
We have already pointed out that the infinite time life of metastable states 
is an artifact of mean field theory. We therefore do not expect to find
a dynamical transition at $T_d > T_c$ in a short range system. On the 
other hand it should be possible to observe a ``reminiscence'' of the mean 
field $T_d > T_c$ in the 1RSB case by looking at the energy density 
relaxation when the system, starting from a random configuration, is quenched 
abruptly (i.e. infinite cooling rate) to a low $T$ value. 

Recent numerical results on a glass-forming system \cite{copa,copari} 
seem to show that $T_d$ is identifiable in short-range models as the 
temperature below which the energy density $e(t)$ displays a two step 
relaxation. The first step corresponds to the relaxation of the system to 
some metastable states with a mean field like behaviour \cite{frpa}, 
$e(t) \propto t^{-\alpha}$ where $\alpha$ is weakly depending on $T$. The 
second one, that happens on a remarkably larger time scale, is due to the 
slow decay of metastable states by activated processes.

We study the relaxation behaviour of $e(t)$ by considering a non small
size $L$=25. Starting from a random initial configuration the system is
quenched to the final temperature and it evolves by Monte Carlo, 
each spin being suggested to flip in one MC step. Presented data are 
averaged over two independent dynamics.

\begin{figure}[htbp]
\begin{center}
\leavevmode
{\epsfig{figure=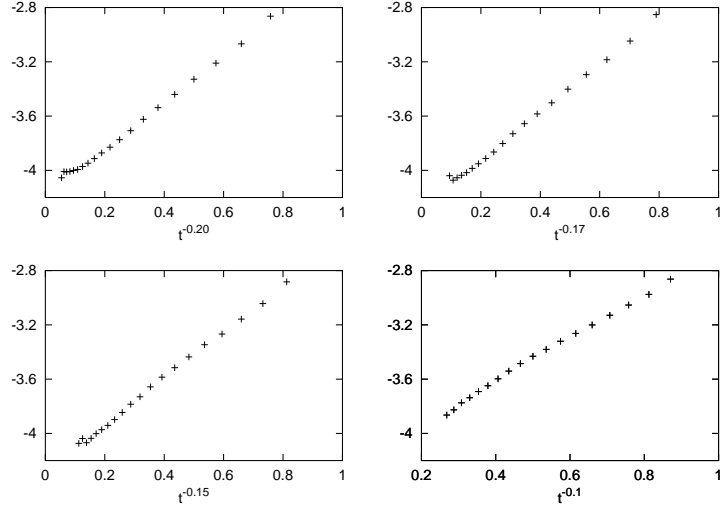,angle=270,width=10cm}} 
\caption{Data on $e$ as a function of $y=t^{-d_e/z(T)}$ at different 
temperatures. From left to right and top to bottom,
$T$=2.6(+) , $2.3 (\times)$, $1.95 (\ast)$ and $1.3 (\Box)$, where
$y$=$t^{-0.20}$, $t^{-0.17}$, $t^{-0.15}$ and $t^{-0.10}$ respectively.}
\end{center}
\label{enescala}
\end{figure}

The energy density $e$ is presented in [Fig. 8] as a 
function of $y$=$t^{-\alpha_e(T)}$ at the different temperatures considered
$T$=2.60 ($\simeq T_c$), 2.30, 1.95 and 1.30.
These data are well consistent with a linear behaviour of $e$ as a function
of $y$ also at large times (small $y$), i.e. there is no evidence for a 
two step relaxation. Moreover at the higher temperatures we have considered 
the system seems near to the equilibrium at the end of the time window and 
correspondingly the values of $e$ result compatible with the equilibrium ones 
obtained from the previously shown data on smaller systems.

The observed behaviour gives evidence for the dynamical critical temperature 
coinciding with the statical one, $T_d=T_c$, as 
it happens in models with FRSB. 

Since we are looking at a second order transition, $\alpha$
is related to the dynamical critical exponent $z$ by $\alpha_e(T)=d_e/z(T)$,
where $d_e = d - 1/ \nu$. We get $\alpha_e \simeq 0.2$ at $T=2.6 \simeq T_c$
that agrees well with the previous estimates \cite{frpa} $\nu \simeq 2/3$, 
$z(T_c) \simeq 7$. Our statistics is inadequate to give meaningful results 
on the behaviour of $z(T)$, which appears however definitely dependent on $T$,
the obtained estimation being consistent with $z(T)=z(T_c) \: T_c/T$.

To conclude this section we present in [Fig. 9] data on
\begin{equation}
C(2t,t)= \frac{1}{N} \sum_{i=1}^{N} \langle \frac{1}{M} 
  \sum_{l=1}^{M} {\sigma_i}^l(2t) {\sigma_i}^l(t) \rangle
\end{equation}
plotted as a function of $y=t^{-{\alpha_q}_c}$, where 
${\alpha_q}_c= \alpha_q(T \simeq T_c) \simeq 0.06$.

At $T=2.6 \simeq T_c$ the observed behaviour is consistent with $C$ 
going to zero linearly in $t^{-{\alpha_q}_c}$. By scaling laws one has 
${\alpha_q}_c=d_q/z(T_c)$, where $d_q=(d-2+ \eta)/2$. We note that  
${\alpha_q}_c \simeq 0.06$ agrees with $\eta \simeq 0$, 
$z(T_c) \simeq 7$. Our statistics results inadequate to fit data at lower
temperatures where the behaviour results consistent with $C$ going to a 
non-zero value for $t \rightarrow \infty$ at $T < T_c$.

\begin{figure}[htbp]
\begin{center}
\leavevmode
\centerline{\epsfig{figure=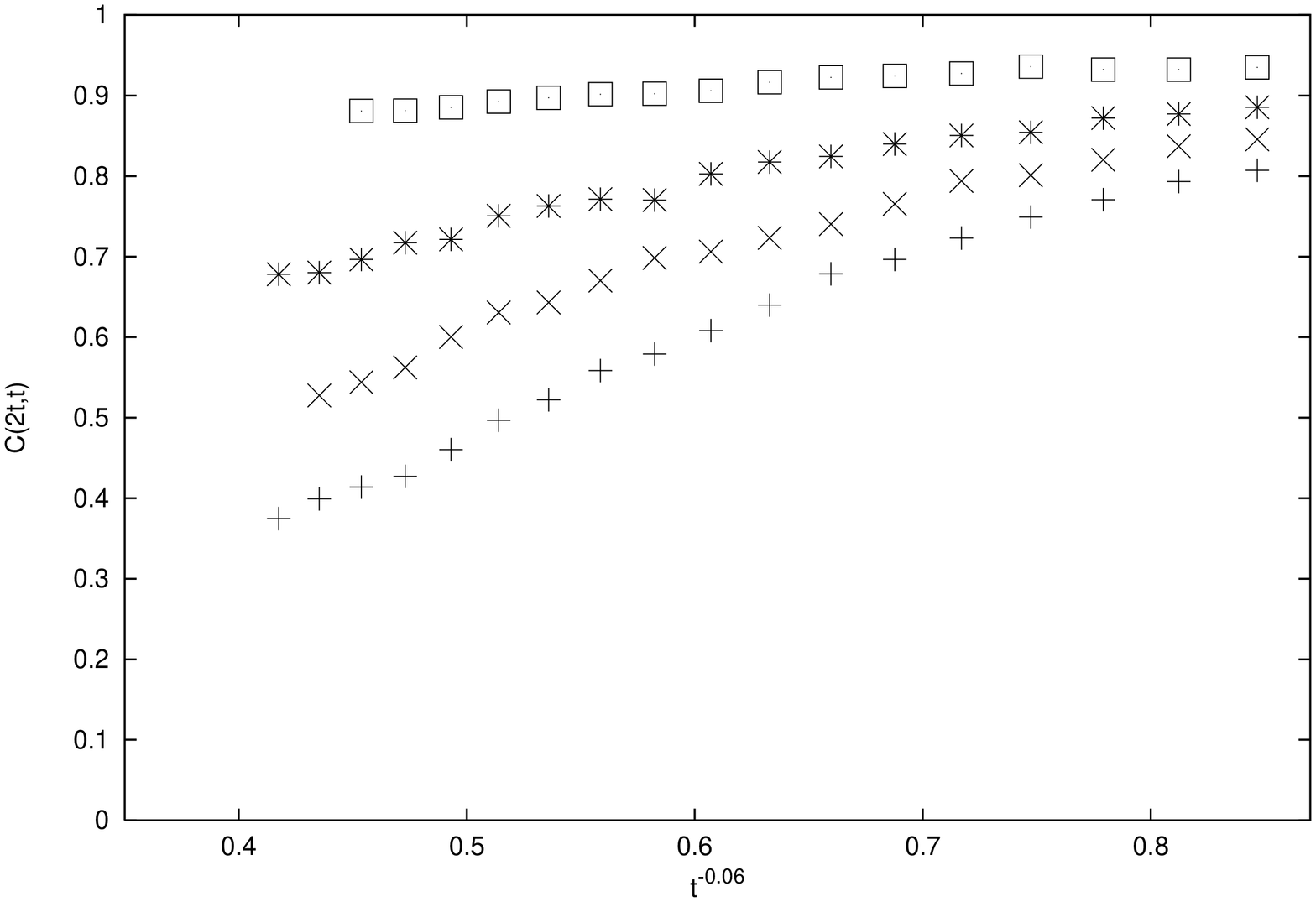,angle=270,width=10cm}} 
\end{center}
\caption{Data on $C(2t,t)$ as a function of $y=t^{-d_q/z(T_c)}=t^{-0.06}$ at
$T$=2.6(+) , $2.3 (\times)$, $1.95 (\ast)$ and $1.3 (\Box)$.}
\label{ctc}
\end{figure}

\subsection{Aging and generalized FDT}
\noindent
The data presented in this section are obtained by MC simulations on
a $L$=16 system at $T$=2.0, i.e. well below the critical temperature. 
We average over 50 different dynamics (initial conditions and 
realizations of the thermal noise). Starting from a random configuration, the 
system is quenched abruptly to the final temperature. After $t_w$ steps we 
apply a small perturbing magnetic field $\delta h=h_0=0.4$, measuring the 
two times correlation function
\begin{equation}
C(t+t_w,t_w)= \langle \frac{1}{M} \sum_{l=1}^{M} {\sigma_i}^l(t+t_w)
{\sigma_i}^l (t_w) \rangle
\end{equation}
and the staggered magnetization
\begin{equation}
m_s[h](t+t_w)= \frac{1}{N} \sum_{i=1}^{N} \langle 
\frac{1}{M} \sum_{l=1}^{M} \sigma_i^l(t+t_w) \rangle.
\end {equation}

The behaviour of $C(t+t_w,t_w)$ for different values of $t_w$ is shown in 
[Fig. 10] on the left. On the right we plot the same data
as function of $t/t_w$. The hypothesis of simple aging, i.e. 
$C(t+t_w,t_w) = \tilde{C}(t/t_w)$, seems to be not well verified.

\begin{figure}[htbp]
\begin{center}
\leavevmode
\epsfig{figure=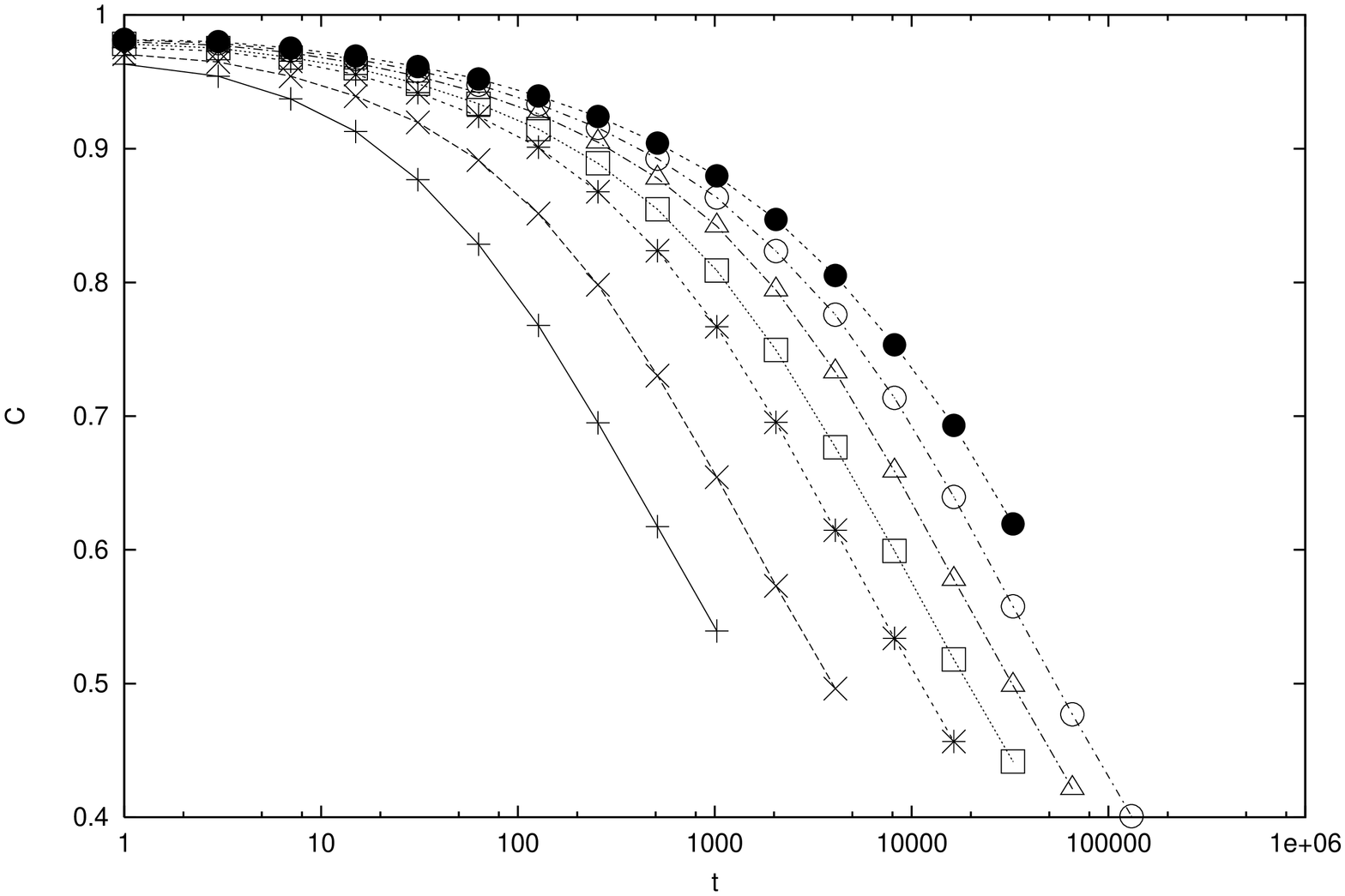,angle=270,width=6cm}
\epsfig{figure=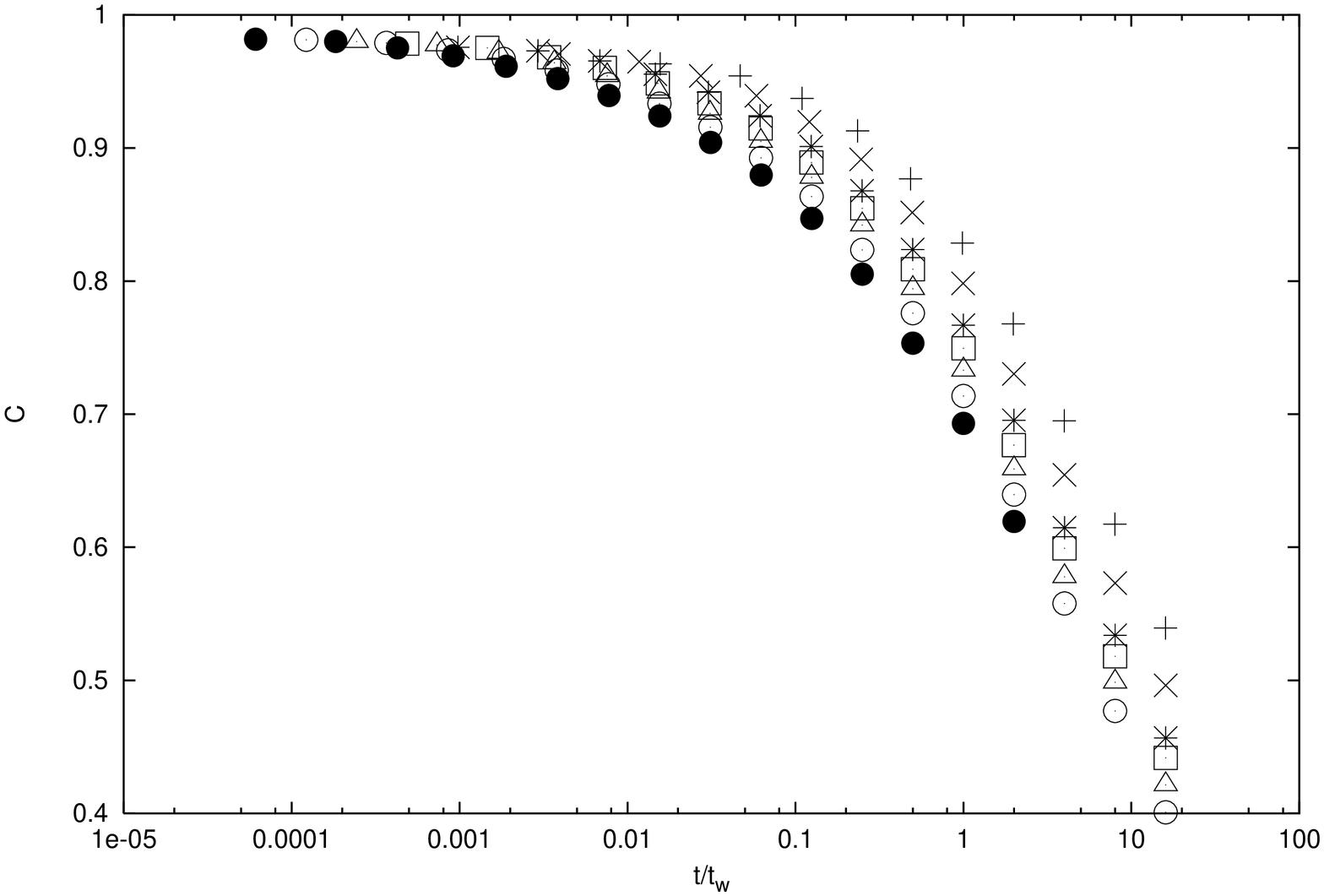,angle=270,width=6cm}
\caption{Data on $C(t+t_w,t_w)$ plotted as function of $t$ (left) and as 
function of $t/t_w$ (right) for $t_w$=$2^6(+)$, $2^8 (\times)$, 
$2^{10} (\ast)$, 
$2^{11} (\Box)$, $2^{12}(\triangle)$, $2^{13}(\bigcirc)$, $2^{14} (\bullet)$.}
\end{center}
\label{cscala}
\end{figure}

In [Fig. 11] we plot $m_s[h](t+t_w)/(\beta \: h_0)$ as a 
function of $C(t+t_w,t_w)$ for different values of $t_w$. Here the expected 
scaling is quite well satisfied and we are therefore allowed to take the 
large times behaviour of this 
quantity as an estimation for $S(C)$. The behaviour of
$m_s[h](t+t_w,t_w)/(\beta \: h_0)$ at $t_w=2^{17}$ (i.e. the largest value
considered) is presented in [Fig. 12] together with the estimation
of $y(q)= \int_q^1 dq' \int_{0}^{q'} dq'' P(q'')$ as obtained by the $L$=5
data on $P(q)$. As expected, the dynamical function $S(C)$ results perfectly 
compatible with the statical one $y(q)$. This dynamical quantity, 
obtained by simulating a non small system, gives further evidence
that the model does not behave like MF 1RSB spin glasses.

\begin{figure}[htbp]
\begin{center}
\leavevmode
\centerline{\epsfig{figure=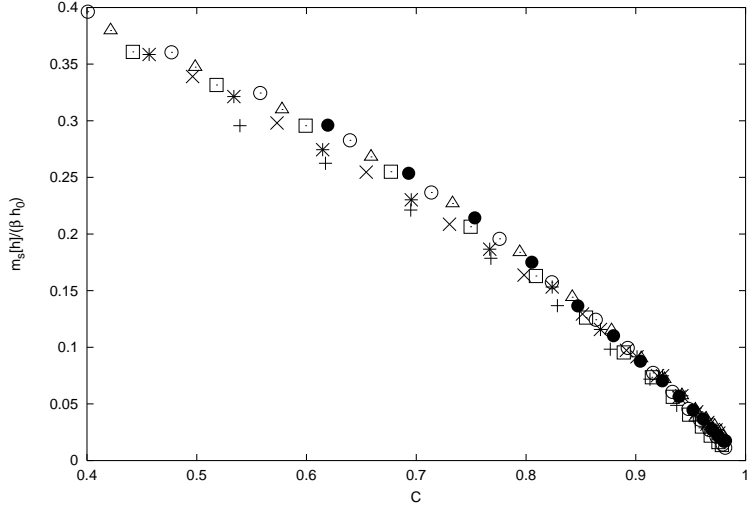,angle=270,width=10cm}} 
\caption{Data on $m_s[h](t+t_w)/(\beta \: h_0)$ plotted as function of 
$C(t+t_w,t_w)$ 
for $t_w$=$2^6(+)$, $2^8 (\times)$, $2^{10} (\ast)$, 
$2^{11} (\Box)$, $2^{12}(\triangle)$, $2^{13}(\bigcirc)$, $2^{14} (\bullet)$.}
\end{center}
\label{crscala}
\end{figure}

\begin{figure}[htbp]
\begin{center}
\leavevmode
\centerline{\epsfig{figure=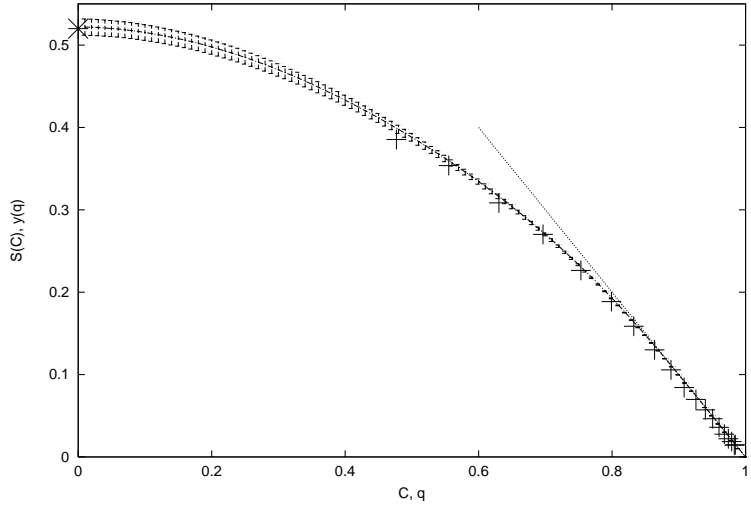,angle=270,width=10cm}} 
\caption{Data obtained from the dynamics on 
$m_s[h](t+t_w)/(\beta \: h_0)$ plotted as function of $C(t+t_w,t_w)$ at 
$t_w$=$2^{17}(+)$ compared with the 
behaviour of $y(q)$ as obtained from the $P(q)$ for $L$=5 (the line with
error bars). 
The $\ast$ represents a dynamical estimation for $\lim_{C \rightarrow 0} 
S(C)$ as obtained by a simulation in which the magnetic field is switched 
on from the beginning. We also plot 1-C (see the text).}
\end{center}
\label{yyy}
\end{figure}

We have shown in the figure the line corresponding to $1-C$ 
just to emphasize the small $t$ region where FDT is
satisfied. The $q_1$ value is evaluable as the one where $y$ begins to move 
away from this line. In mean field 1RSB models we observe a straight line 
for $y$ going from $q_1$ to $0$ (the dynamical function $S(C)$ showing the
same structure than the statical one):
\begin{equation}
y(q) = \left \{
\begin{array}{ccc} 
(1-m) (q_1-q) + (1-q_1) & \hspace{.3in} \mbox{for} \hspace{.3in} & 
q \leq q_1 \\
1-q & \hspace{.3in} \mbox{for} \hspace{.3in} & q \ge q_1 \\
\end{array}
\right.
\label{scsc}
\end{equation}
This behaviour has been recently observed in simulations of structural
glasses \cite{pa4}, by defining appropriately the correlation function $C$
and the response function to small perturbations $R$.

In the case of our model, instead of been in agreement with (\ref{scsc}), 
$S(C)$ (we dynamically estimate $S(0)$ extrapolating data obtained by a 
simulation in which $h$ has been switched on from the beginning) results more 
similar to the one observable in FRSB models \cite{frari,mapariru}, confirming 
the results on the equilibrium probability distribution $P(q)$ at low 
temperatures.

\section{Discussion and conclusions}

Summing up our main results:
\begin{itemize}
\item The transition appears to be of second order, characterized by
a diverging correlation length for $T \rightarrow T_c \simeq 2.6$. Our 
estimations for the critical exponents agree with the ones previously
obtained \cite{frpa} $\nu \simeq 2/3$, $\eta \simeq 0$, $z \simeq 7$.

\item There is no evidence for the dynamical critical temperature $T_d$
being greater then the statical one,  
where in finite dimensional models that are 1RSB in mean field $T_d > T_c$
should mark the onset of two steps relaxation processes, .

\item The distribution probability of the overlaps $P(q)$ results
non trivial at low temperature but its behaviour is very different from
the one of mean field 1RSB models.

\item Well below $T_c$, we find a $P(q)$ still definitely non-zero on the 
whole interval $[-q_1,q_1]$. This result is confirmed by out of equilibrium
simulations on a quite larger lattice (i.e. by the behaviour of $S(C)$ which
is related to $P(q)$ by the generalized FDT relation).

\item The behaviour of $P(q)$ seems nevertheless also not compatible
with the one usually encountered in FRSB models, this resulting well
evident if one looks at the cumulant $g(T)$. 
\end{itemize}

Our possible interpretation of this phenomenology is the following. 
In this model the disorder is not space-invariant. In each particular 
sample there will be a finite density of regions which are more likely than 
the rest of the system to freeze into a SG phase. Hence when $T >T_c$ there is
a non-zero probability to find a region of space where the system is
 in the 1RSB phase.
The typical size of these regions will diverge when $T$ approaches $T_c$ 
since the system freezes at the critical temperature.
In these regions the spins are very correlated and one can formally divide
the total high temperature spin glass susceptibility into the contribution of 
bubbles 
of frozen spins and the contribution of the rest of the system.  
We write
\be
\chi_{SG} = \chi_{B}+ \chi_{R} 
\ee
where $B$ stands for ``bubbles''  and $R$ for ``rest''.
So, for $T$ approaching $T_c$ from above one has

\be 
\chi_{B} \propto (T-T_c)^{-\gamma} \hspace{1cm}  \chi_{R} = O(1) 
\ee

It is important to note that this kind of transition, 
at least from this analytical point of view, 
is very different from an ordinary second order transition in mean field, 
where a zero mass 
mode {\em on the mean field solution} causes diverging correlations. 
Here the high temperature solution is stable and non-perturbative 
effects dominate the approach to the transition. 
Within the bubbles the transition can still be similar to the MF 1RSB.
It is the weight of the bubbles that grows continuously and diverges
 when $T \to T_c$.

It is possible that a similar phenomenon is responsible for the fact
that the function $P(q)$ does not tend to the sum of two delta functions 
at $T \lt T_c$ (opposite to what happens in mean field theory). This point
deserves a more careful study.

\section*{Acknowledgments}
We are happy to thank Felix Ritort for a careful reading of the manuscript
and for fruitful related discussions.

\end{document}